\documentclass[conference]{IEEEtran}

\usepackage{graphicx}
\usepackage{textcomp}
\usepackage{xcolor}
\usepackage{color}
\usepackage{graphicx}
\usepackage{url}
\usepackage{cite}
\usepackage{amsthm}
\usepackage{enumerate}
\usepackage{tabularx}
\usepackage{booktabs}
\usepackage{algorithm}
\usepackage{algpseudocode}
\usepackage{multirow}
\usepackage[normalem]{ulem}
\usepackage{tikz}
\usetikzlibrary{shapes,arrows}
\usepackage{amsmath}
\usepackage{amssymb}
\usepackage{colortbl}
\usepackage{xspace}
\usepackage{graphics}
\usepackage{bm}
\usepackage{enumitem}
\usepackage{soul}
\sethlcolor{yellow}
\usepackage{bigstrut}
\usepackage{adjustbox}
\usepackage{comment}
\usepackage{graphicx}
\usepackage{subcaption}

\IEEEoverridecommandlockouts
\makeatletter
\DeclareRobustCommand*\cal{\@fontswitch\relax\mathcal}
\makeatother

\begin{document}
\bstctlcite{IEEEexample:BSTcontrol}
\title{Link Quality Aware Pathfinding for Chiplet Interconnects}

\author{\IEEEauthorblockN{Aaron Yen\textsuperscript{*}}
\IEEEauthorblockA{\textit{ECE Department} \\
\textit{University of California, Los Angeles}\\
Los Angeles, USA \\
aaronhyen@ucla.edu}
\and
\IEEEauthorblockN{Jooyeon Jeong\textsuperscript{*}}
\IEEEauthorblockA{\textit{ECE Department} \\
\textit{University of California, Los Angeles}\\
Los Angeles, USA \\
jooyeon37@ucla.edu}
\and
\IEEEauthorblockN{Puneet Gupta}
\IEEEauthorblockA{\textit{ECE Department} \\
\textit{University of California, Los Angeles}\\
Los Angeles, USA \\
puneetg@ucla.edu}
\thanks{\textsuperscript{*}Aaron Yen and Jooyeon Jeong contributed equally to this work.}
}

\maketitle
\vspace{-5pt}

\begin{abstract}

As chiplet-based integration advances, designers must select among short-reach die-to-die interconnect technologies with widely varying shoreline and areal bandwidth density, energy per bit, reach, and raw bit error rate (BER). Meeting stringent delivered BER targets in chiplet systems requires error-correcting codes (ECC), but incurs energy, area, and throughput overheads. We develop a flow centered around RTL synthesis power and area estimations to support pathfinding of inter-chiplet links under a stringent $10^{-27}$ delivered BER target. We synthesize a parameterized Reed-Solomon code with CRC-64 and Go-Back-N retry logic to estimate the correction overhead for different transceiver bit error rates.

Results show that ECC can materially change link comparisons under common figures of merit and that CRC+ARQ can reduce the required RS strength (and decoder overhead) at moderate BERs while still meeting stringent delivered-BER targets. 
We present a CP-SAT-based link assignment formulation that uses these ECC-corrected metrics under reach, delivered-bandwidth, and shoreline constraints in system-level optimization.

\footnote{Accepted at IEEE Electronic Components and Technology Conference 2026.}
\end{abstract}

\section{Introduction}
Chiplet-based integration improves yield and enables heterogeneous composition, but shifts many bottlenecks onto die-to-die interconnects. Advanced packaging supports various interconnects, creating a broad design space with wide ranges of (i) shoreline bandwidth density (Gbps/mm perimeter), (ii) areal bandwidth density (Gbps/mm\textsuperscript{2} of chip area), (iii) energy/bit (pJ/bit), (iv) reach, and (v) raw bit error rate (BER). While emerging die-to-die standards like UCIe~\cite{ucie_spec}, AIB~\cite{intel_aib_spec}, and BoW~\cite{ocp_bow_spec} exist, system-level requirements span short-reach (<1,mm) on-package links to longer links across large interposers and wafer-scale packages.

Prior work on advanced packaging proposed interconnect metrics like shoreline bandwidth density and signaling figures of merit for heterogeneous integration~\cite{Jangam2020-wm,Jangam2021-kj}. However, despite strict BER and latency requirements for intra-package communication, these metrics for chiplet interconnects do not yet account for ECC-related power, area, and throughput overheads needed for reliable communication.

To meet strict delivered-BER targets, links often combine forward error correction (FEC) with a cyclic redundancy check (CRC) and automatic repeat request (ARQ). Early-stage link comparisons, however, typically focus only on nominal physical metrics and ignore error-correction overhead. As link energy reaches the sub-pJ/bit regime, ECC logic can consume a significant fraction of total link energy and area, altering comparisons under common figures of merit \cite{Jangam2020-wm}.

To achieve high reliability with low latency variability (and thus smaller retransmission buffers), we focus on Reed-Solomon FEC (RS-FEC)~\cite{ReedSolomon1960}, which provides strong burst-error correction and predictable performance under harsh channels. Standard treatments discuss FEC/ARQ tradeoffs and hybrid schemes~\cite{LinCostello}, and prior work studied energy–reliability tradeoffs between correction and retransmission for on-chip communication~\cite{demichel2003errorcontrol,bertozzi2005errorcontrol}.

For CRC-based detection, we use CRC-64/ECMA-182~\cite{ecma182} for its low undetected-error probability, which depends on message length, polynomial, and error structure. Koopman~\cite{koopman2004crc} and the FAA integrity report~\cite{faa_crc_report} provide guidance beyond the uniform $2^{-k}$ approximation. At the system level, chiplet placement and interconnect co-optimization have been studied in prior work~\cite{yenai2020}. This paper makes contributions:
\begin{itemize}[leftmargin=*]
\item A synthesis-based ECC characterization flow for RS-FEC, CRC64, and Go-Back-N retry that quantifies energy, area, and throughput overhead as a function of raw BER under a delivered-BER target.
\item A link metric calculator that produces ECC-corrected throughput, energy, and area metrics suitable for system-level optimization.
\item A CP-SAT \cite{ortools,cpsatlp2023}-based link assignment tool that maps link technologies to chiplet nets under reach, bandwidth, and shoreline constraints.
\end{itemize}

\section{Motivation}

\begin{figure*}[htbp]
    \centering
    \includegraphics[width=1.6\columnwidth]{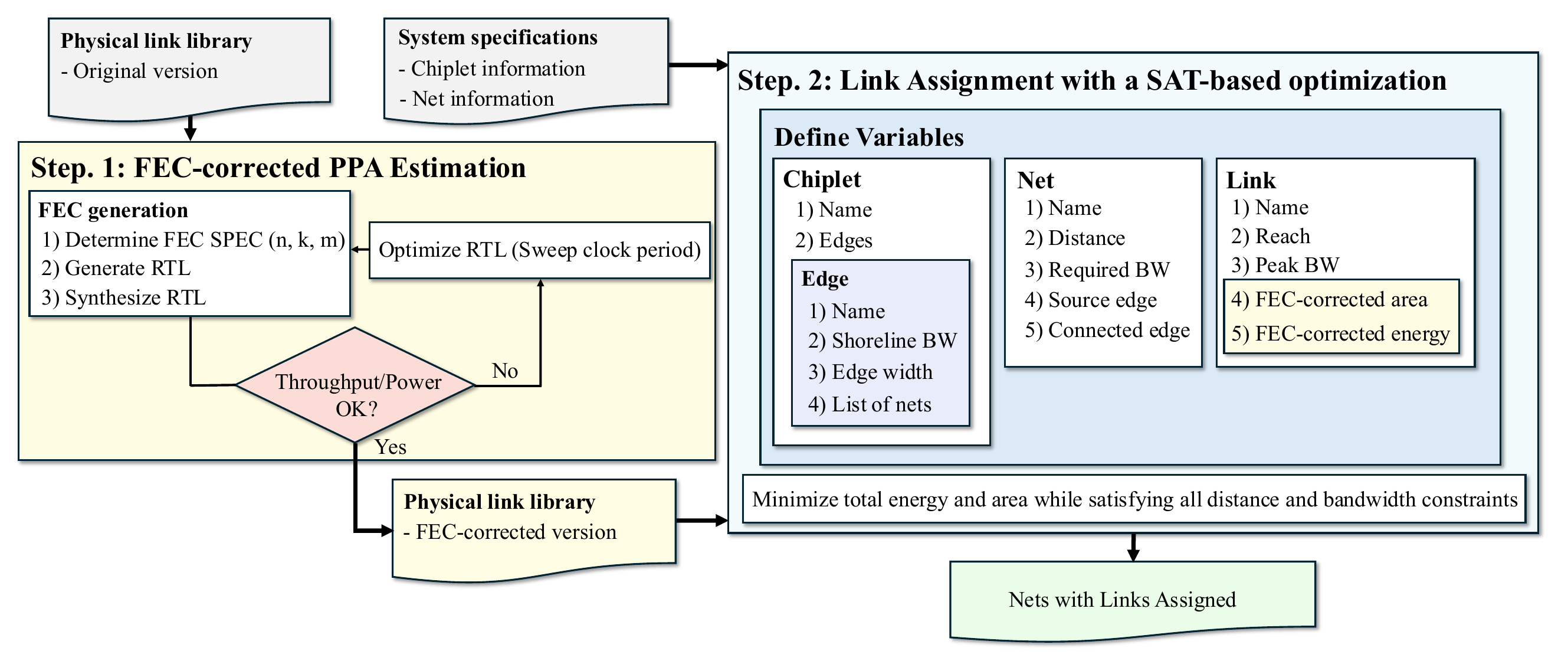}
    \caption{Overall flow of link assignment.}
    \label{flow}
\end{figure*}

Chiplet link pathfinding often uses uncorrected transceiver metrics (e.g., energy/bit, bandwidth density) and reach constraints. In practice, many systems need delivered BER far below raw BER across operating conditions. Bridging this gap requires ECC, which adds three coupled costs: (1) \emph{code-rate overhead} reducing goodput, (2) \emph{energy overhead} from encoder/decoder logic and retry control, and (3) \emph{area/throughput overhead} requiring multiple ECC blocks at high line rates.

We distinguish \emph{post-FEC BER} (bit error rate after RS decoding, before CRC) from \emph{delivered BER} (payload-bit error rate after CRC and ARQ). FEC-only targets post-FEC BER directly (and thus delivered BER in the absence of retry), while the hybrid FEC+CRC+ARQ scheme targets delivered BER by bounding CRC-undetected residual errors.

These costs are nonlinear with BER. As shown in Section~\ref{sec:results}, the RS code strength required to meet a fixed delivered-BER target increases rapidly as input BER degrades, which increases decoder energy and area. Protocol choices also matter: CRC64 and retry can relax the required RS strength at moderate BERs while still meeting a delivered-BER target.

\begin{figure}[!t]
    \centering
    \includegraphics[width=\columnwidth]{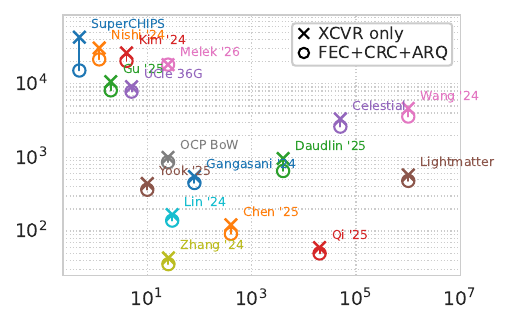}
    \caption{FoM vs. reach (mm) for representative links. Crosses denote raw transceiver FoM, and open circles show ECC-corrected FoM using hybrid FEC+CRC+ARQ. FoM is defined as payload throughput density (Gbps/mm) divided by energy per delivered bit (pJ/bit).}
    \label{fig:reach_vs_fom}
    \vspace{-8.6mm}
\end{figure}

Pathfinding decisions based only on raw link metrics can be misleading. Fair comparison requires incorporating BER-driven ECC overhead, and system-level link assignment must consider these corrected metrics along with physical layout constraints such as shoreline bandwidth and reach. Figure~\ref{fig:reach_vs_fom} illustrates how ECC correction can shift a common link figure of merit across representative reach regimes.\footnote{Our modeling and assignment scripts are open sourced at \url{https://github.com/nanocad-lab/chiplet-link-pathfinding}}

\section{Methodology}
\label{sec:method}

Figure~\ref{flow} illustrates the overall workflow. The link metric calculator first characterizes each interconnect technology in terms of ECC-corrected metrics, which are then combined with net-to-edge mapping and physical distances. Finally, the CP-SAT optimization module assigns link technologies to nets.

\subsection{Link Protection Stack and Metrics}
We model a standard ``FEC corrects, CRC detects, retry recovers'' link protection stack: payload and protocol header bytes are protected by a cyclic redundancy check, then Reed-Solomon forward error correction parity is applied over the (header+payload+CRC) bytes. We use CRC64/ECMA-182~\cite{ecma182} and RS codes~\cite{ReedSolomon1960,LinCostello} over GF$(2^8)$. At the receiver, RS-FEC attempts to correct symbol errors, CRC64 detects remaining corruption, and a Go-Back-N (GBN) automatic repeat request mechanism retransmits frames that fail CRC.

We report two sets of link-level metrics: (1) \emph{FEC-only}, where RS-FEC is sized to meet the delivered BER target without relying on retry and (2) \emph{FEC+CRC+ARQ}, where CRC64 and retry are used to meet the delivered BER target and RS-FEC is only used as needed to keep the retry rate acceptably low. In the FEC-only mode we omit CRC and retry, so RS protects only header+payload bytes. In the FEC+CRC+ARQ mode, RS protects header+payload+CRC bytes.

\subsection{Code Selection and Reliability Model}
To evaluate the impact of bit errors on interconnects, we synthesize byte-oriented RS-FEC codes. We fix the codeword length to $N=86$ symbols over GF$(2^8)$, motivated by PCIe 6.0 FLIT-mode interleaving ($\approx$85\,B groups)~\cite{pcie6_flit_fec}, and vary the number of information symbols $K$ such that the code can correct $t = \lfloor \frac{N-K}{2} \rfloor$ symbol errors per codeword.

For \emph{FEC-only}, we select the highest-rate (largest $K$) RS$(86,K)$ code whose estimated post-FEC BER meets a target post-FEC BER. Without retry, post-FEC BER equals the delivered BER. Assuming i.i.d. symbol errors, the number of symbol errors per codeword $X$ is $X\sim\mathrm{Binomial}(N,p_\mathrm{sym})$, with $p_\mathrm{sym}=1-(1-p_\mathrm{pre})^M$. Post-FEC BER is approximated as the expected fraction of erroneous bits in uncorrectable codewords:
\begin{equation}
    \mathrm{BER}_\mathrm{post} \approx \sum_{i=t+1}^{N}\left(\frac{1}{2}\cdot\frac{i}{N}\right)\Pr[X=i],
\end{equation}
where we assume random bit corruption within each erroneous symbol (byte).

For \emph{FEC+CRC+ARQ}, we account for two reliability failure modes: (1) \emph{silent data corruption} (SDC) when CRC fails to detect a residual error, and (2) \emph{drop} when a frame exceeds a bounded number of CRC-detected retransmissions. As a simplifying collision model, we approximate the CRC undetected-error probability as $p_\mathrm{undet}$. Unless otherwise noted we use $p_\mathrm{undet}=2^{-64}$ under random residual error patterns. Prior analyses show the undetected-error probability depends on message length, polynomial choice, and error structure~\cite{koopman2004crc,faa_crc_report}. We apply a conservative ``catastrophic'' corruption model: if an undetected error is delivered, a fraction $f_\mathrm{wrong}$ of payload bits are incorrect.

Let $p_\mathrm{frame\_fail}$ denote the per-attempt probability that an ARQ frame still contains corruption after RS decoding (and would therefore fail CRC absent a miss). A CRC-detected failure occurs with probability
\begin{equation}
    p_\mathrm{det} \approx p_\mathrm{frame\_fail}\cdot(1-p_\mathrm{undet}).
\end{equation}
Under independent attempts and retry-on-detect, the probability that a \emph{delivered} frame is undetected is $p_\mathrm{frame\_fail}\cdot p_\mathrm{undet}/(1-p_\mathrm{det})$, giving
\begin{equation}
    \mathrm{BER}_\mathrm{delivered} \approx f_\mathrm{wrong}\cdot \frac{p_\mathrm{frame\_fail}\cdot p_\mathrm{undet}}{1-p_\mathrm{det}}.
\end{equation}

To bound retries, we cap retransmissions to at most $R$ (so the total number of attempts is $A=R+1$). The frame-drop probability is then
\begin{equation}
    p_\mathrm{drop} \approx p_\mathrm{det}^{A}.
\end{equation}
To map drops into a bit-level budget compatible with $\mathrm{BER}_\mathrm{target}$, we use a 1-bit-equivalent model $\mathrm{BER}_\mathrm{drop,eff} = p_\mathrm{drop}/(8P)$, where $P$ is payload bytes.

Given $\mathrm{BER}_\mathrm{target}$, we derive a frame-fail budget $p_\mathrm{frame\_target}$ as the tighter of the SDC and drop constraints, and select the highest-rate RS$(86,K)$ such that $p_\mathrm{frame\_fail}\le p_\mathrm{frame\_target}$. Unless otherwise noted, we use $\mathrm{BER}_\mathrm{target}=10^{-27}$ (UCIe-motivated~\cite{ucie_spec}), $f_\mathrm{wrong}=0.5$, and $R=1$ (one retry) for the FEC+CRC+ARQ mode.

\subsection{Frame-Fail Probability and Goodput}
We treat RS-FEC as a \emph{streaming} byte code: parity overhead is attributed proportionally by code rate ($K/N$) without per-frame RS padding waste. Let $P$ be payload bytes, $H$ header bytes, and $C$ CRC bytes (8 for CRC64). We use $P=256$\,B unless otherwise noted (PCIe 6.0 FLIT size~\cite{pcie6_flit_fec}). The RS-protected data bytes per ARQ frame are:
\begin{equation}
    D = P + H + C.
\end{equation}
Given a raw (pre-FEC) bit error probability $p_\mathrm{pre}$, we approximate the symbol error probability as
\begin{equation}
    p_\mathrm{sym} = 1-(1-p_\mathrm{pre})^M.
\end{equation}
This symbol-error mapping and the binomial tail model below assume independent bit errors. 
We model the probability of codeword decode failure as a binomial tail over symbol errors:
\begin{equation}
    p_\mathrm{blk\_fail} = \Pr[X>t], \quad X \sim \mathrm{Binomial}(N, p_\mathrm{sym}).
\end{equation}
Under the streaming model, an ARQ frame spans an \emph{effective} number of codewords $B_\mathrm{eff}=D/K$ (which can be fractional). We then approximate the frame-fail probability as:
\begin{equation}
    p_\mathrm{frame\_fail} \approx 1-(1-p_\mathrm{blk\_fail})^{B_\mathrm{eff}}.
\end{equation}
Fractional $B_\mathrm{eff}$ corresponds to cross-frame streaming RS coding and ignores per-frame padding overhead. The wire bytes transmitted per attempt under the streaming code-rate model are:
\begin{equation}
    \mathrm{wire\_bytes/attempt} \approx D\cdot \frac{N}{K}.
\end{equation}
Assuming independent attempts and retry-on-detect, the expected number of attempts per delivered frame is $\mathbb{E}[\mathrm{attempts}] \approx 1/(1-p_\mathrm{det})$ and the goodput is:
\begin{equation}
    \mathrm{goodput} = \frac{P}{(\mathrm{wire\_bytes/attempt})\cdot \mathbb{E}[\mathrm{attempts}]}.
\end{equation}
This retry model approximates per-frame independent retransmission. We synthesize a Go-Back-N controller with a replay window sized to the bandwidth--delay product, but in analysis we approximate retransmission cost using a geometric attempts factor and ignore go-back-N pipeline flushes or window-full stalls, which are negligible in our rare-retry target regime~\cite{Hayashida_1993}, but higher retry rates or larger RTT/BDP may require a BDP-aware ARQ throughput model.

\subsection{Synthesis}
Using the code selection method above, we choose a worst-case input pre-FEC BER of $10^{-3}$ and a post-FEC BER target of $10^{-27}$, which sets the worst-case code to RS$(86,44)$. We synthesize RS$(86,K)$ codes from RS$(86,44)$ to RS$(86,84)$ in steps of $t=1$. We used Synopsys Design Compiler~\cite{DC} with the ASAP7 7nm predictive PDK~\cite{asap7_pdk}, which is the latest open standard-cell library \cite{asap_1, asap_3, asap_4, asap_5}. We use a fixed clock period that meets timing for the worst-case code (0.8\,ns).

In addition to RS-FEC, we synthesize CRC64 append/check blocks and a parameterized GBN retry controller. The CRC/retry sweep spans payload size and round-trip time (RTT), which determines the retry window depth and replay buffering. Reported GBN area/power includes both controller logic and the synthesized replay buffer (implemented using SRAM macros when available and inferred storage otherwise).

To compare different interconnect links, we normalize ECC area and energy by effective throughput. RS-FEC throughput is decoder-limited at the synthesized clock, while CRC/retry assumes single-frame-per-cycle at 500\,MHz. Dynamic power is estimated with a vectorless activity model (randomized data, clock gating enabled, leakage excluded), so pJ/bit values should be interpreted as estimates. For RS-FEC receiver energy, we approximate correction activity by weighting incremental decoder energy beyond syndrome by $p_\mathrm{corr}=\Pr[1\le X\le t]$ under the i.i.d. symbol-error model.

We treat chiplet transceivers as largely mixed-signal blocks and do not scale published transceiver metrics across nodes. We synthesize ECC logic in ASAP7 (7\,nm) and scale ECC energy and area to a 3\,nm using public TSMC guidance.

\subsection{Chiplet Link Assignment}
We formulate a constrained optimization problem that maps logical interconnect nets to physical link technologies while minimizing system-level cost. Each net must be assigned exactly one link type that satisfies its reach and bandwidth requirements, and the aggregate shoreline width consumed on each chiplet edge must not exceed the available physical I/O width. In the chiplet, the available physical I/O width depends on which edge the connection is made to, as the required routing distance varies accordingly. We model the I/O-available length as half of the chiplet perimeter. 

We solve this problem using the OR-Tools constraint programming satisfiability (CP-SAT) solver~\cite{ortools,cpsatlp2023}. Table~\ref{tab:variables_cpsat} summarizes the notation. For each net $n \in \mathcal{N}$ and link type $l \in \mathcal{L}$, the binary variable $x_{n,l}$ indicates whether $n$ uses link type $l$. The shoreline bandwidth density ${BW}^{shore}_l$, energy (${Energy}_l$), and areal bandwidth density (${BW}^{area}_l$) parameters incorporate ECC overhead, so that the assignment problem operates on delivered payload bandwidth and energy. We normalize the objective using the total power and total area ($\lambda_P$ and $\lambda_A$).

\begin{table}[htbp]
    \centering
    \caption{Notation for CP-SAT-based I/O link assignment.}
    \footnotesize
    \begin{tabularx}{\columnwidth}{
      >{\hsize=0.4\hsize}X
      >{\hsize=1.6\hsize}X
    }
        \toprule
        \multicolumn{2}{l}{\textbf{Input parameters}} \\ 
        \midrule
        $\mathcal{C,L,N}$ & Set of chiplets, available link types and nets \\
        $\mathcal{N/W}_{c,e}$ & Net/Edge width of chiplet $c$ edge $e$ \\
        $d_n$ & Actual physical distance required for net $n$ [mm] \\
        $r_l$ & Maximum reach distance supported by link type $l$ [mm] \\
        ${Energy}_l$ & Energy per bit for link type $l$ [pJ/bit] \\        
        ${BW}^{{shore}}_l$ & Shoreline bandwidth density for link type $l$ [Gbps/mm]\\
        ${BW}^{{req}}_n$ & Required bandwidth for net $n$ [Gbps]\\
        ${BW}^{{area}}_l$ & Areal bandwidth density for link type $l$ [Gbps/mm$^2$] \\
        $\lambda_{P/A}$ & Total system power/area of chiplets/wafer [W, mm$^2$] \\
        \midrule
        \multicolumn{2}{l}{\textbf{Decision variables}} \\
        \midrule
        $x_{n,l}$ & 1 if net $n$ uses link type $l$, 0 otherwise \\
        $w_{n,l}$ & Shoreline width allocated to net $n$ if it uses link type $l$ \\
        \bottomrule
\end{tabularx}
\label{tab:variables_cpsat}
\end{table}

The objective minimizes a weighted sum of total link power and total transceiver area:
\begin{equation}
    \label{eq:cp-sat-objective}
    \begin{aligned}
        \min \quad & \sum_{n \in \mathcal{N}} \sum_{l \in \mathcal{L}}
        \bigl(
            ({Energy}_l \cdot {BW}^{{req}}_n) / \lambda_P  \cdot x_{n,l} \\
            & \qquad\quad + ({BW}^{{req}}_n / {BW}^{{area}}_l)/ \lambda_A \cdot x_{n,l}
        \bigr).
    \end{aligned}
\end{equation}

\begin{algorithm}[htbp]
\footnotesize
\caption{Chiplet I/O link assignment via CP-SAT}
\label{alg:cp-sat}
\begin{algorithmic}[1]
    \State \textbf{Initialize model:} $\mathcal{M} \gets \text{CpModel()}$
\ForAll{nets $n \in \mathcal{N}$ and link types $l \in \mathcal{L}$}
    \State Define binary variable $x_{n,l} \in \{0,1\}$
    \State Define width variable $w_{n,l} \in [0, W_\mathrm{max}]$ 
    \State \textbf{Constraint 1}: Reachability
    \If{$d_n > r_l$}
        \State $x_{n,l} = 0$
    \EndIf
    \State \textbf{Constraint 2}: Bandwidth capacity
    \State $w_{n,l} \cdot {BW}^{{shore}}_l \ge {BW}^{{req}}_n \, x_{n,l}$    
\EndFor
\ForAll{chiplets $c \in \mathcal{C}$ and edges $e \in \mathcal{E}$}
\State \textbf{Constraint 3}: Edge capacity limit
\State $\sum_{n \in \mathcal{N}_{c,e}} \sum_{l \in \mathcal{L}} w_{n,l} \le W_{c,e}$ 
\EndFor
\ForAll{nets $n \in \mathcal{N}$}
    \State \textbf{Constraint 4}: Single link per net
    \State $\sum_{l \in \mathcal{L}} x_{n,l} = 1$
\EndFor
    \State \textbf{Objective}: Minimize Eq.~\eqref{eq:cp-sat-objective}
    \State \textbf{Solve model} $\mathcal{M}$ with CP-SAT solver
    \State \textbf{Return:} Assigned link variables $x_{n,l}$ and $w_{n,l}$
\end{algorithmic}
\end{algorithm}

\section{Experimental Results}
\label{sec:results}

\subsection{RS-FEC Energy, Rate, and Area Trends}
\begin{figure}[!t]
	\centering
	\includegraphics[width=\columnwidth]{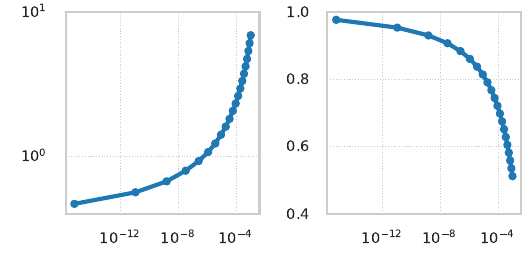}
	\caption{Selected RS-FEC decoder energy per info bit and code rate vs input BER for a post-FEC BER target of $10^{-27}$ (ASAP7 synthesis sweep, highest-rate $K$ meeting target). Left: energy per info bit [pJ/bit]. Right: code rate ($K/N$).}
	\label{fig:rs_side_by_side}
\end{figure}

Figure~\ref{fig:rs_side_by_side} shows how RS-FEC decoder energy per information bit and effective code rate vary with input BER for RS$(86,K)$ codes when selecting the highest-rate code that meets a post-FEC BER target of $10^{-27}$. As BER degrades, stronger codes (smaller $K$) are required, reducing code rate and increasing decoder energy.



\begin{figure}[!t]
	\centering
	\includegraphics[width=\columnwidth]{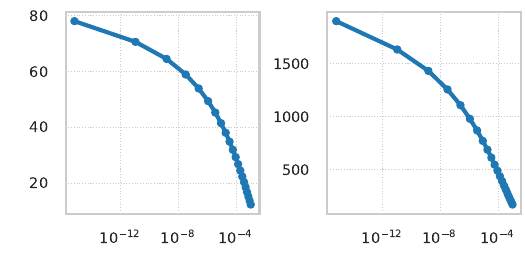}
    \caption{RS-FEC throughput density vs. input BER (ASAP7, post-FEC BER = $10^{-27}$). X-axis: pre-FEC BER. Left: throughput density (Gbps/mm). Right: throughput density (Gbps/mm$^2$).}
	\label{fig:rs_density}
\end{figure}

Figure~\ref{fig:rs_density} reports RS-FEC information-throughput density vs input BER for a post-FEC BER target of $10^{-27}$ (ASAP7, with decoder-limited throughput), normalized to both a 1-D (Gbps/mm) and 2-D (Gbps/mm$^2$) footprint. The Gbps/mm metric uses a square-footprint proxy (width $=\sqrt{\mathrm{area}}$).

\subsection{FEC-only vs FEC+CRC+ARQ Tradeoffs}
\begin{figure}[!t]
	\centering
	\includegraphics[width=\columnwidth]{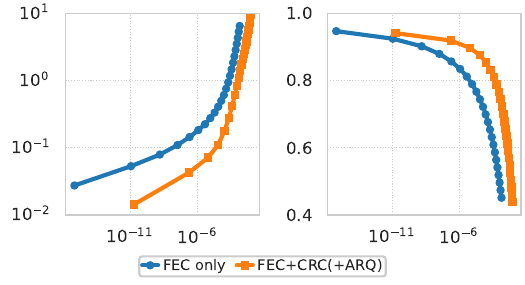}
	\caption{ECC energy and goodput vs input BER (ASAP7), comparing FEC-only and a no-drop (unbounded-retry) FEC+CRC+ARQ hybrid. Left: energy per payload bit [pJ/bit]. Right: goodput rate.}
	\label{fig:asap7_side_by_side}
\vspace{-5mm}
\end{figure}

Figure~\ref{fig:asap7_side_by_side} compares two protection modes for a representative frame size ($P$=256\,B payload, $H$=8\,B header, CRC64) under a delivered BER target of $10^{-27}$: \emph{FEC-only} (no retry) and \emph{FEC+CRC+ARQ} (retry on CRC fail with RS-FEC sized to meet the retry budget). The FEC+CRC+ARQ scheme relaxes the required RS strength at moderate BERs, improving goodput and reducing ECC energy while still meeting the delivered BER target. At very low input BER, it can fall back to a no-FEC operating point ($K=N$).

For example, at input BER $\approx 10^{-4}$, FEC-only selects RS$(86,62)$. Under an unbounded-retry (no-drop) hybrid model, FEC+CRC+ARQ selects RS$(86,78)$, reducing ECC energy from $\approx 0.61$ to $\approx 0.18$\,pJ/payload-bit and improving goodput from $\approx 0.70$ to $\approx 0.85$. Enforcing a one-retry limit ($R{=}1$) and budgeting drops into the same $10^{-27}$ target instead selects RS$(86,72)$ (ECC energy $\approx 0.31$\,pJ/payload-bit, goodput $\approx 0.79$).



To ensure numerical stability for rare RS decode-failure probabilities, binomial tails are computed via log-domain summation (instead of $1-\mathrm{CDF}$ subtraction) and verified against high-precision references for representative RS$(86,K)$ points. Relative error remains below $10^{-12}$ for probabilities down to $\sim10^{-27}$, validating the computation under the i.i.d. error assumption.


\subsection{CRC64 and Go-Back-N Retry Synthesis}
\begin{figure}[!t]
    \centering
	    \begin{subfigure}[t]{\columnwidth}
	        \centering
	        \includegraphics[width=\linewidth]{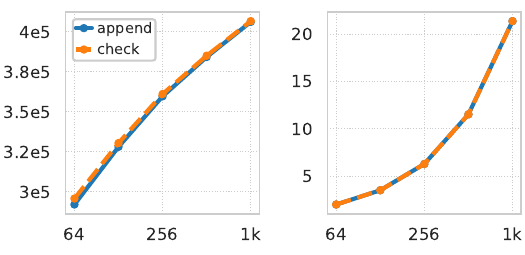}
	        \caption{CRC64 append/check (left: throughput density [Gbps/mm$^2$] vs payload bytes, right: dynamic power [mW] vs payload bytes).}
	        \label{fig:crc_area_power}
	    \end{subfigure}
	    \begin{subfigure}[t]{\columnwidth}
	        \centering
	        \includegraphics[width=\linewidth]{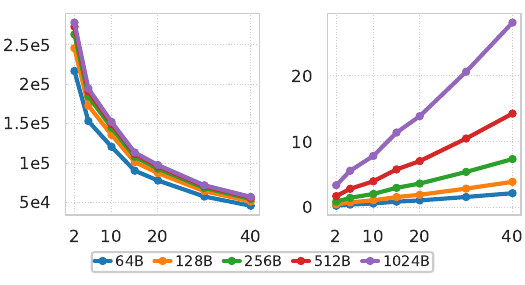}
	        \caption{GBN retry controller (left: throughput density [Gbps/mm$^2$] vs RTT [ns], right: dynamic power [mW] vs RTT [ns]).}
	        \label{fig:gbn_area_power_rtt}
    \end{subfigure}
    \caption{CRC64 and GBN retry synthesis sweeps (ASAP7, 500\,MHz, vectorless activity model).}
    \label{fig:crc_gbn_sweeps}
\end{figure}

We synthesize CRC64 append/check and a parameterized GBN retry controller across payload size and RTT. The synthesized GBN block includes the replay buffer memory sized to the outstanding window implied by the bandwidth--delay product (BDP). Reported area/power includes this buffer (SRAM macros when available, inferred storage otherwise). For the baseline ($P$=256\,B, $H$=8\,B, CRC64, RTT=10\,ns at 500\,MHz), the window is 7 frames, corresponding to $\approx$1.9\,KB of replay storage. Figure~\ref{fig:crc_gbn_sweeps} shows representative ASAP7 sweeps.

Table~\ref{tab:crc_retry_cost} reports a concrete operating point (256\,B payload, 8\,B header, 10\,ns RTT) showing that CRC64 and retry logic have small per-bit energy and area overhead relative to RS-FEC at moderate-to-high BERs. Energy per payload bit is computed from dynamic power as $E = (P_\mathrm{dyn}/f_\mathrm{clk})/(8P)$ at $f_\mathrm{clk}=500$\,MHz (dynamic power only, leakage excluded). Power uses a vectorless activity model with randomized data inputs (static probability 0.5, toggle rate 0.5 toggles/cycle) and non-stalling control (valid/ready asserted).

\begin{table}[!t]
    \centering
    \caption{CRC64 and GBN retry synthesis cost at 256\,B payload, 8\,B header, 10\,ns RTT (ASAP7, 500\,MHz).}
    \scriptsize
    \setlength{\tabcolsep}{3pt}
    \begin{tabular}{lrrr}
        \toprule
        \textbf{Block} & \textbf{Area} [$\mu$m$^2$] & \textbf{Dyn. Pwr.} [mW] & \textbf{Energy} [pJ/payload-bit] \\
        \midrule
        CRC append & 2847 & 6.28 & 0.00614 \\
        CRC check  & 2836 & 6.28 & 0.00614 \\
        GBN retry  & 7071 & 2.05 & 0.00201 \\
        \bottomrule
    \end{tabular}
    \label{tab:crc_retry_cost}
\end{table}

\subsection{Link Assignment with Various Configurations}

\begin{table*}[t]
    \centering
    \caption{Final link metrics comparison for representative links, using ASAP7-synthesized ECC (7\,nm) and a 3\,nm ECC scaling case study. Transceiver energy and bandwidth metrics are taken as published and are not scaled across nodes. Due to space limitations, $Gbps/mm$ and $Gbps/$$\mathrm{mm}^2$ are abbreviated as $G/mm$ and $G/$$\mathrm{mm}^2$ in the table.}
    \label{link-summary}
    \setlength{\tabcolsep}{2.2pt}
    \begin{scriptsize}
        \begin{tabular}{l|cccc|ccc|ccc|ccc|ccc}
            \toprule
	            \multirow{2}{*}{\textbf{Link Name}} & \multirow{2}{*}{\textbf{\begin{tabular}[c]{@{}c@{}}Reach\\ (mm)\end{tabular}}} & \multirow{2}{*}{\textbf{\begin{tabular}[c]{@{}c@{}}Proc.\\ (nm)\end{tabular}}} & \multirow{2}{*}{\textbf{\begin{tabular}[c]{@{}c@{}}Raw\\ BER\end{tabular}}} & \multirow{2}{*}{\textbf{\begin{tabular}[c]{@{}c@{}}Link\\ Type\end{tabular}}} & \multicolumn{3}{c|}{\textbf{7nm (FEC Only)}} & \multicolumn{3}{c|}{\textbf{7nm (FEC+CRC)}} & \multicolumn{3}{c|}{\textbf{3nm (FEC Only)}} & \multicolumn{3}{c}{\textbf{3nm (FEC+CRC)}} \\
            \cmidrule(lr){6-8} \cmidrule(lr){9-11} \cmidrule(lr){12-14} \cmidrule(lr){15-17}
            & & & & & $G/mm$ & $G/mm^2$ & $pJ/b$ & $G/mm$ & $G/mm^2$ & $pJ/b$ & $G/mm$ & $G/mm^2$ & $pJ/b$ & $G/mm$ & $G/mm^2$ & $pJ/b$ \\
            \midrule
	            SuperCHIPS \cite{shih2021superchips} & 0.5 & 16 & 1e-14 & E & 1144 & 1608 & 0.08 & 1103 & 1719 & 0.07 & 1144 & 4856 & 0.05 & 1103 & 5172 & 0.05 \\
	            Hsu '21 \cite{hsu2021ultra} & 1.0 & 7 & 1e-25 & E & 5187 & 1020 & 0.50 & 4998 & 2063 & 0.50 & 5187 & 1605 & 0.48 & 4998 & 2100 & 0.50 \\
	            Nishi '23 \cite{nishi2023shortreach} & 1.2 & 5 & 1e-25 & E & 10744 & 1420 & 0.33 & 10353 & 5053 & 0.33 & 10744 & 2885 & 0.32 & 10353 & 5280 & 0.32 \\
	            Nishi '24 \cite{nishi2024clocked} & 1.2 & 5 & 1e-12 & E & 5530 & 1043 & 0.25 & 5332 & 1074 & 0.25 & 5530 & 1843 & 0.22 & 5332 & 1843 & 0.23 \\
	            Kang '25 \cite{kang2025energy} & 1.5 & 28 & 1e-12 & E & 17163 & 486 & 1.21 & 16547 & 483 & 1.24 & 17163 & 610 & 1.18 & 16547 & 595 & 1.22 \\
	            Gu '25 \cite{gu2025highspeed} & 2.0 & 3 & 1e-15 & E & 3661 & 1442 & 0.43 & 3530 & 1525 & 0.43 & 3661 & 3605 & 0.40 & 3530 & 3739 & 0.41 \\
	            Kim '25/26 \cite{kim2026nextgen} & 2.0 & 28 & 1e-12 & E & 8734 & 1560 & 0.34 & 8421 & 1662 & 0.34 & 8734 & 4442 & 0.31 & 8421 & 4690 & 0.32 \\
	            Wang '25 \cite{wang2025ultrahigh} & 3.0 & 28 & 1e-16 & E & 10256 & 1189 & 1.27 & 9882 & 2944 & 1.30 & 10256 & 2069 & 1.25 & 9882 & 3019 & 1.29 \\
	            Kim '24 \cite{kim2024} & 4.0 & 28 & 1e-12 & E & 11442 & 1526 & 0.53 & 11031 & 1623 & 0.54 & 11442 & 4181 & 0.51 & 11031 & 4390 & 0.52 \\
	            GLink 2.3LL \cite{guc2025glink} & 5.0 & 5 & 1e-20 & E & 13990 & 1345 & 0.33 & 13480 & 4189 & 0.33 & 13990 & 2591 & 0.32 & 13480 & 4343 & 0.33 \\
	            UCIe 36G \cite{guc2025ucie} & 5.0 & 5 & 1e-20 & E & 3647 & 699 & 0.45 & 3514 & 1050 & 0.45 & 3647 & 931 & 0.43 & 3514 & 1059 & 0.44 \\
	            Yook '25 \cite{yook2025lowlatency} & 10 & 4 & 1e-15 & E & 641 & 552 & 1.65 & 618 & 550 & 1.70 & 641 & 716 & 1.62 & 618 & 700 & 1.67 \\
	            Melek '26 \cite{melek2026terabit} & 25 & 3 & 1e-27 & E & 5270 & 4216 & 0.29 & 5270 & 4216 & 0.29 & 5270 & 4216 & 0.29 & 5270 & 4216 & 0.29 \\
	            OCP BoW \cite{ocp2021bow} & 25 & 12 & 1e-20 & E & 488 & 1564 & 0.54 & 471 & 7662 & 0.55 & 488 & 3549 & 0.52 & 471 & 8194 & 0.54 \\
	            Vandersand '25 \cite{vandersand2025scaling} & 25 & 3 & 1e-27 & E & 448 & 393 & 0.52 & 448 & 393 & 0.52 & 448 & 393 & 0.52 & 448 & 393 & 0.52 \\
	            Zhang '24 \cite{zhang2024areaefficient} & 25.4 & 28 & 1e-12 & E & 59 & 505 & 1.54 & 56 & 502 & 1.59 & 59 & 639 & 1.51 & 56 & 624 & 1.57 \\
	            Lin '24 \cite{lin2024subpJ} & 30 & 28 & 1e-12 & E & 344 & 355 & 2.32 & 332 & 350 & 2.39 & 344 & 416 & 2.29 & 332 & 405 & 2.37 \\
	            Poon '21 \cite{poon2021monolithic} & 30 & 7 & 1e-12 & E & 830 & 496 & 1.35 & 800 & 493 & 1.39 & 830 & 625 & 1.33 & 800 & 610 & 1.37 \\
	            Gangasani '24 \cite{gangasani202456gbps} & 80 & 5 & 1e-9 & E & 564 & 358 & 1.26 & 557 & 372 & 1.24 & 564 & 432 & 1.22 & 557 & 435 & 1.22 \\
	            Chen '25 \cite{chen2025optical} & 400 & 4 & 2.5e-6 & E & 539 & 186 & 6.55 & 564 & 209 & 6.15 & 539 & 218 & 6.44 & 564 & 234 & 6.10 \\
	            Daudlin '25 \cite{daudlin2025nature} & 4e3 & 28 & 6e-8 & O & 278 & 897 & 0.51 & 283 & 1146 & 0.43 & 278 & 1994 & 0.44 & 283 & 2370 & 0.40 \\
	            Qi '25 \cite{qi2025longreach} & 2e4 & 45 & 1e-12 & O & 162 & 220 & 3.04 & 156 & 215 & 3.14 & 162 & 243 & 3.01 & 156 & 235 & 3.12 \\
	            Celestial \cite{celestial2026interconnect} & 5e4 & 4 & 1e-8 & O & 7256 & 527 & 2.75 & 7179 & 563 & 2.74 & 7256 & 737 & 2.70 & 7179 & 754 & 2.71 \\
	            Wang '24 \cite{wang2024codesigned} & 1e6 & 28 & 1e-10 & O & 1860 & 1111 & 0.54 & 1839 & 1289 & 0.51 & 1860 & 2348 & 0.50 & 1839 & 2579 & 0.49 \\
	            Lightmatter \cite{lightmatter2025hotchips} & 1e6 & 45 & 1e-9 & O & 1395 & 1015 & 2.87 & 1379 & 1161 & 2.87 & 1395 & 1959 & 2.83 & 1379 & 2113 & 2.85 \\
            \bottomrule
        \end{tabular}
    \end{scriptsize}
\end{table*}

\subsubsection{Experimental Setup and Link Characteristics}

Table~\ref{link-summary} summarizes the link library used in our case studies, including original process node, reach (mm), raw BER, and link type (E for electrical, O for optical). We report ECC-corrected delivered throughput density ($Gbps/mm^2$) and energy ($pJ$ per delivered payload bit) using ASAP7-synthesized ECC (7\,nm) and a 3\,nm ECC scaling case study, while leaving transceiver metrics unchanged.

Overall, CRC reduces delivered energy/bit and increases delivered throughput density by allowing weaker RS codes at a fixed delivered-BER target. When the raw BER is already extremely low (e.g., Melek~\cite{melek2026terabit}), FEC-only and FEC+CRC become similar. Long-reach optical links remain higher-energy than short-reach electrical links for chiplet-scale distances.

\begin{figure}[htbp]
\vspace{-7.5mm}
    \centering
    \begin{subfigure}[b]{0.44\columnwidth}
        \centering
        \includegraphics[width=\textwidth]{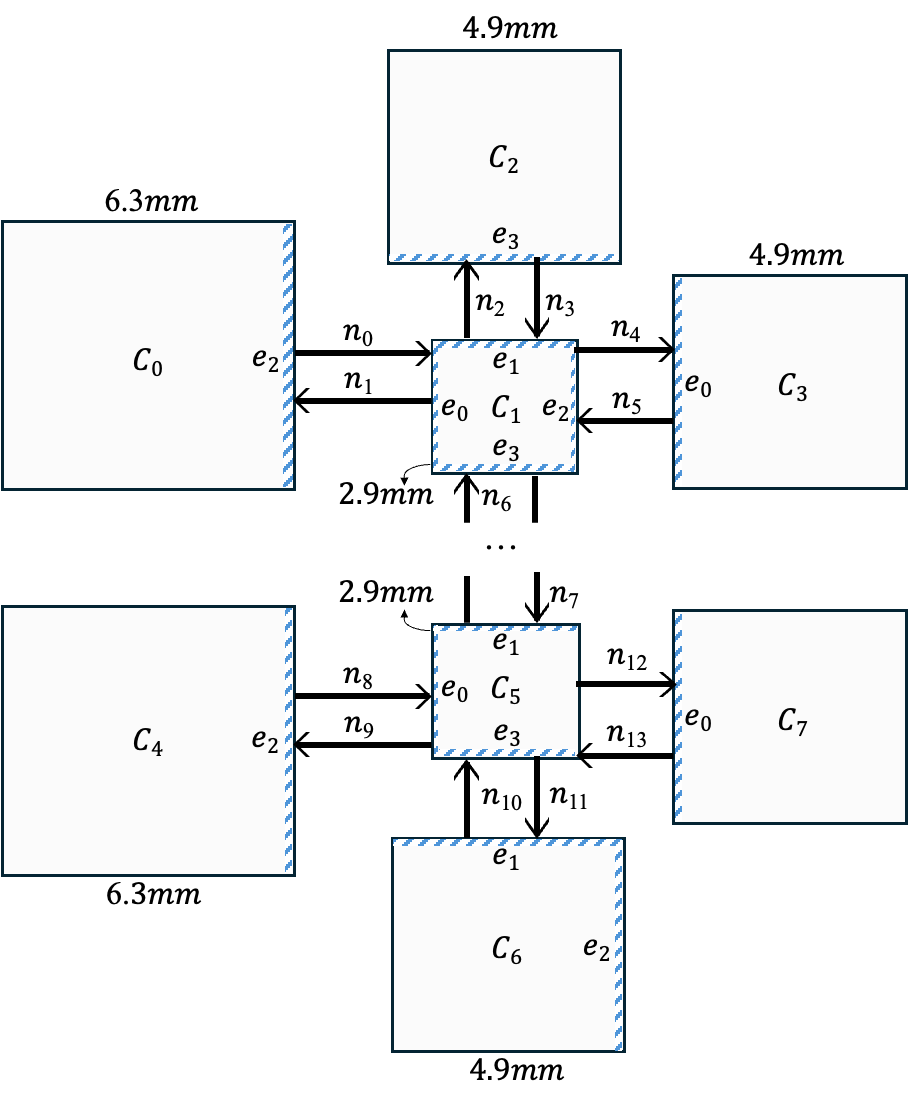}
        \caption{Example 1 \cite{WaferScale}}
        \label{fig:ex1_small}
    \end{subfigure}
    \hfill
    \begin{subfigure}[b]{0.44\columnwidth}
        \centering
        \includegraphics[width=\textwidth]{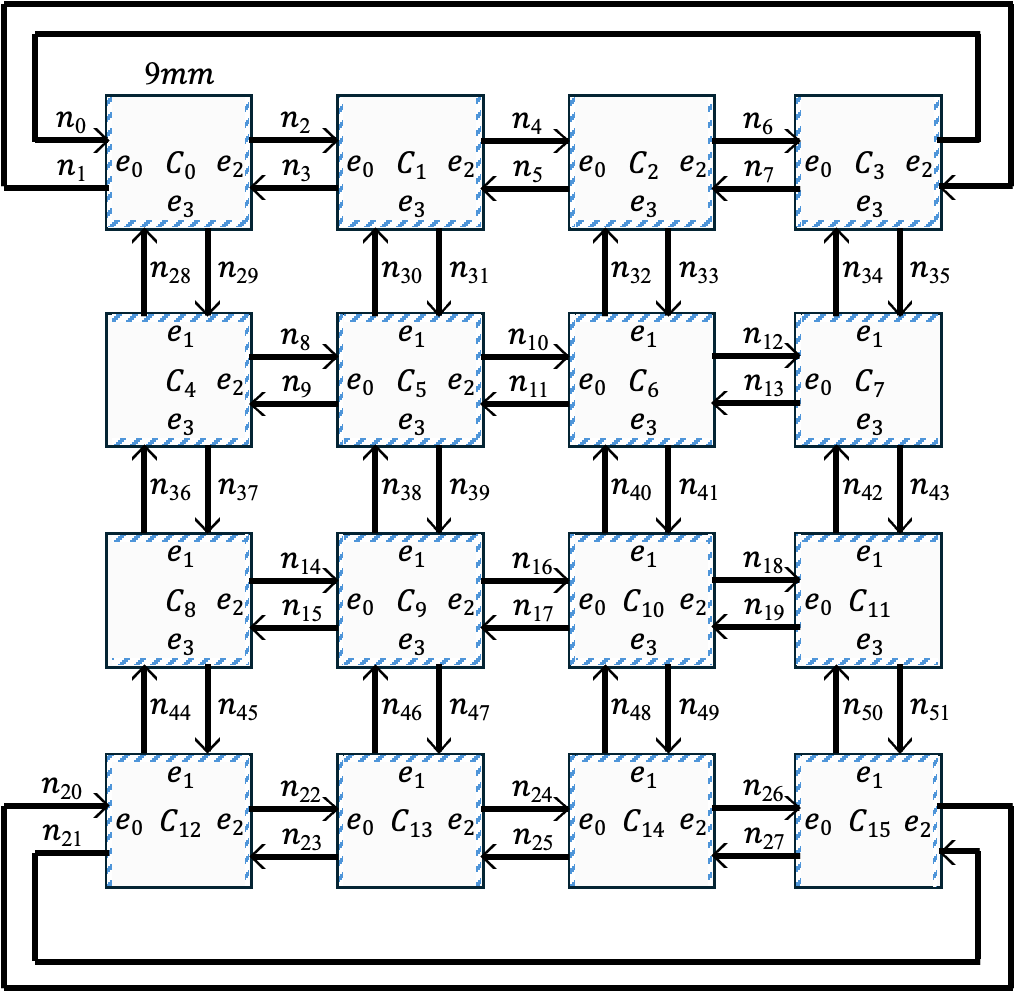}
        \caption{Example 2 \cite{WaferScale}}
        \label{fig:ex2_small}
    \end{subfigure}
    \begin{subfigure}[b]{0.9\columnwidth}
        \centering
        \includegraphics[width=\textwidth]{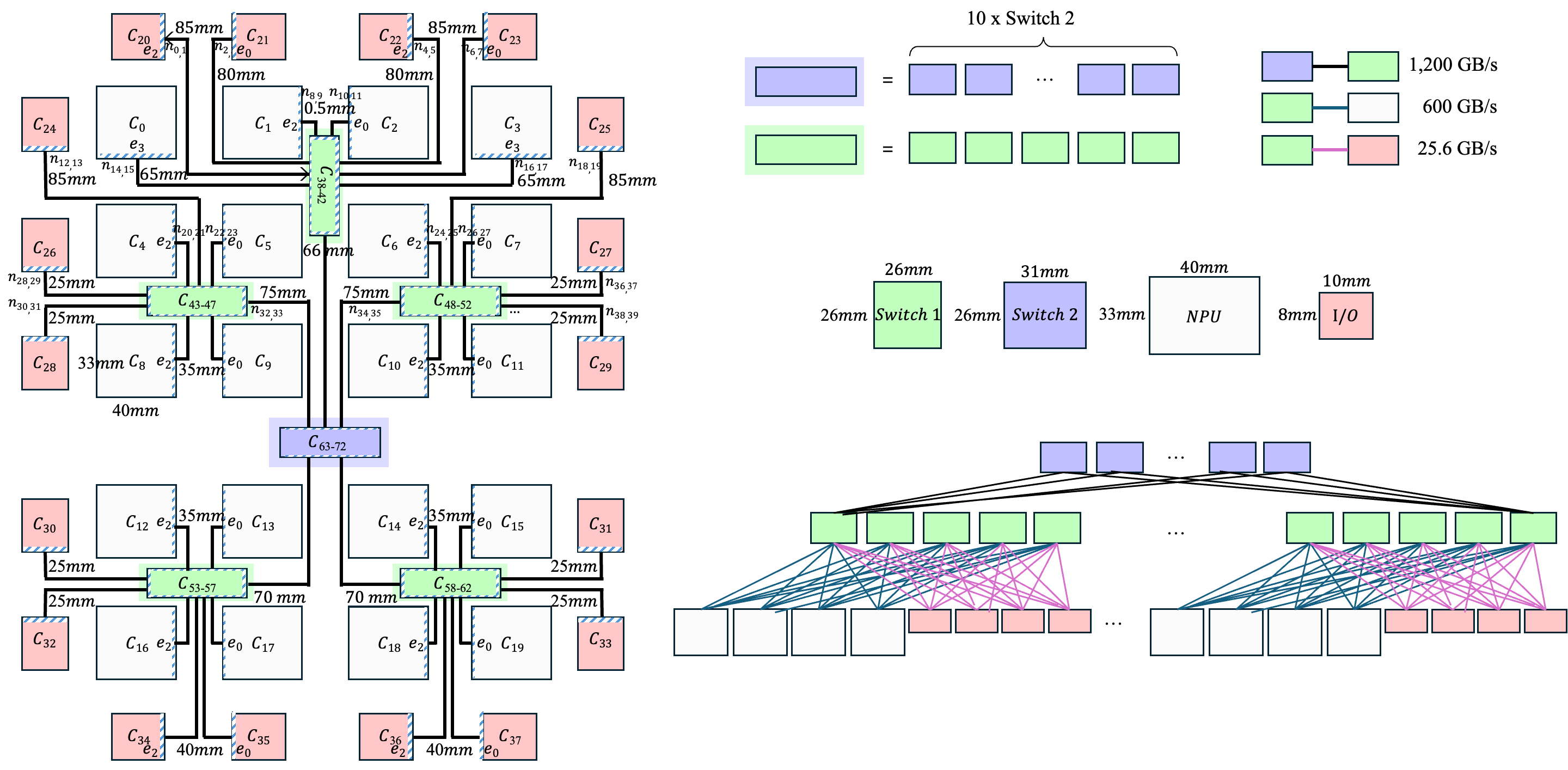}
        \caption{Example 3 \cite{FRED}}
        \label{fig:ex3_large}
    \end{subfigure}
    \caption{Link Assignment Examples 1–3: each chiplet edge has a shoreline budget, and each net connects two edges with a required bandwidth and distance.}
    \label{fig:combined_examples}
\end{figure}

\subsubsection{Evaluated System Configurations}

Figure~\ref{fig:combined_examples} shows three experimental setups. Examples 1 \& 2 come from a wafer-scale architecture~\cite{WaferScale}: Example 1 models a reduced two-tile subset (each tile with one compute die and two memory dies for more net diversity), and Example 2 extends to a $4\times4$ tile array with extra inter-row connections to study link selection in more complex topologies. Example 3 is from prior work~\cite{FRED}, where die dimensions are unspecified, so dimensions are heuristically set to match the reported total area.

Example~1, Example~2, and Example~3 assume total areas/powers of 
175.42~mm$^2$/11.27~W, 
1,403.36~mm$^2$/90.12~W, and 
70,000~mm$^2$/15,000~W, respectively. 
The values for Example~1 and Example~2 are linearly scaled from~\cite{WaferScale} 
based on tile count and are therefore approximate.

\subsubsection{Results and Optimization Analysis}

\begin{figure*}[t]
\centering
\includegraphics[width=1.8\columnwidth]{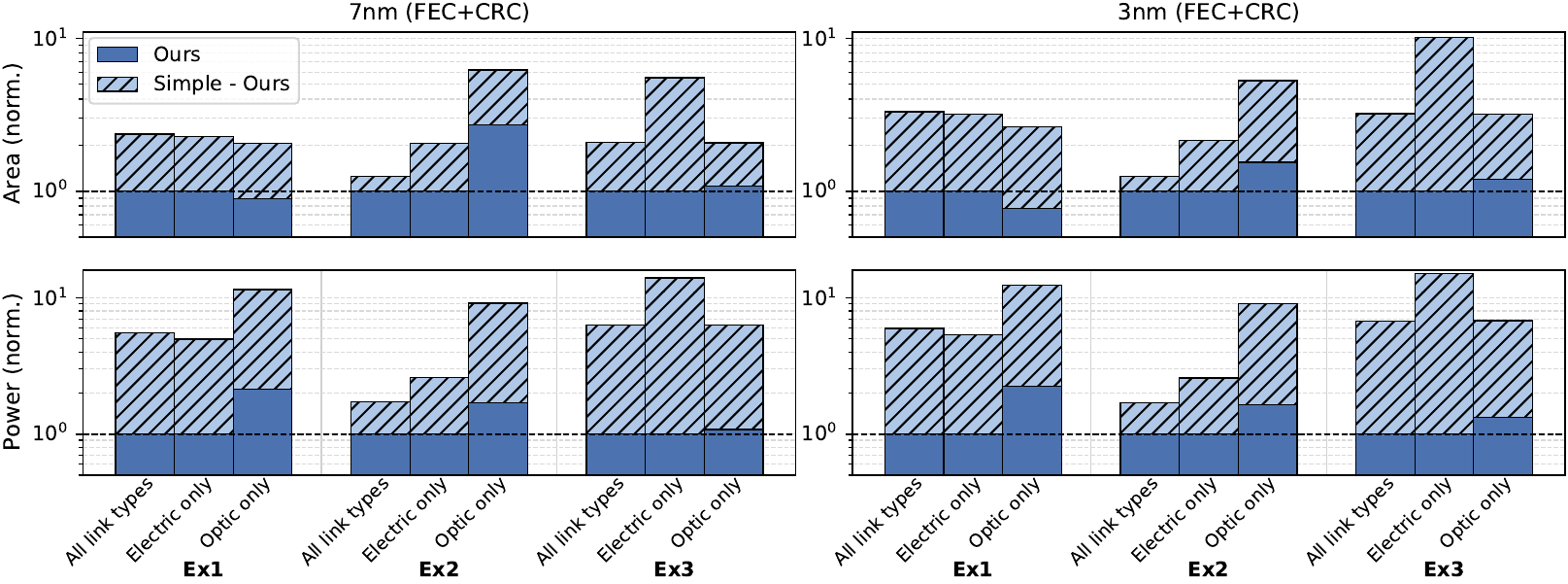}
\caption{Bar-graph summary of FEC+CRC results. Values are normalized per (Example, metric, node) to the ``Ours'' result under All-Link-Types. Bars are stacked: solid = Ours, hatched = (Simple$-$Ours), bar top = Simple.}
\label{fig:full_results_comparison}
\end{figure*}

\begin{table*}[htbp]
    \centering
    \caption{Integrated link selection results with total energy and area comparison for CRC-enabled links scaled to 7nm}
    \label{tab:integrated_results_with_metrics}
     \footnotesize
     \resizebox{\textwidth}{!}{
     \begin{tabular}{llccccc}
        \toprule
        \textbf{Ex. / Case} & \textbf{\# of Nets} & \textbf{Dist. (mm)} & \textbf{BW (Gbps)} & \textbf{All Link Types} & \textbf{Electrical Only} & \textbf{Optical Only} \\
        \midrule
        \multirow{3}{*}{\textbf{Example 1}} 
        & 12 & 0.5 & 2033--7228 & SuperCHIPS \cite{shih2021superchips} / Nishi '24 \cite{nishi2024clocked}& SuperCHIPS \cite{shih2021superchips}/ Nishi '24 \cite{nishi2024clocked}& Wang '24 \cite{wang2024codesigned}\\
        & 2 & 6 & 1604 & Melek '26 \cite{melek2026terabit} & Melek '26 \cite{melek2026terabit} & Wang '24 \cite{wang2024codesigned}\\
        \cmidrule{2-7}
        & \multicolumn{3}{r}{Total Power [W] / Area [mm$^2$]} & 14.07 / 25.55 & 14.07 / 25.55 & 30.15 / 22.82 \\
        \midrule
        \multirow{3}{*}{\textbf{Example 2}} 
        & 48 & 5 & 1604 & Melek '26 \cite{melek2026terabit} & Melek '26 \cite{melek2026terabit} & Daudlin '25 \cite{daudlin2025nature}\\
        & 4 & 60--80 & 1604 & Daudlin '25 \cite{daudlin2025nature}& Gangasani '24 \cite{gangasani202456gbps}& Daudlin '25 \cite{daudlin2025nature} \\
        \cmidrule{2-7}
        & \multicolumn{3}{r}{Total Power [W] / Area [mm$^2$]} & 25.10 / 11.93 & 30.28 / 17.76 & 36.06 / 36.38 \\
        \midrule
        \multirow{5}{*}{\textbf{Example 3}} 
        & 80 & 0.5--25 & 26--600 & Melek '26 \cite{melek2026terabit} & Melek '26 \cite{melek2026terabit} & Daudlin '25 \cite{daudlin2025nature}\\
        & 20 & 0.5--25 & 26--600 & SuperCHIPS \cite{shih2021superchips}& SuperCHIPS \cite{shih2021superchips}& Daudlin '25 \cite{daudlin2025nature}\\
        & 730 & 35--75 & 26--1200 & Daudlin '25 \cite{daudlin2025nature}& Gangasani '24 \cite{gangasani202456gbps}& Daudlin '25 \cite{daudlin2025nature}\\
        & 50 & 35--75 & 26--1200 & Wang '24 \cite{wang2024codesigned}& Gangasani '24 \cite{gangasani202456gbps}& Wang '24 \cite{wang2024codesigned}\\
        \cmidrule{2-7}
        & \multicolumn{3}{r}{Total Power [W] / Area [mm$^2$]} & 313.47 / 310.76 & 881.92 / 958.95 & 318.08 / 313.16 \\
        \midrule
        \multicolumn{7}{c}{\textbf{Case Studies of Example 3}} \\
        \midrule
        \multirow{4}{*}{\textbf{(1) BW $\times$ 1.4}} 
        & 100 & 0.5--25 & 36--840 & Melek '26 \cite{melek2026terabit} / SuperCHIPS \cite{shih2021superchips}& Melek '26 \cite{melek2026terabit} / SuperCHIPS \cite{shih2021superchips}& Daudlin '25 \cite{daudlin2025nature}\\
        & 780 & 35--75 & 36--1680 & Daudlin '25 \cite{daudlin2025nature}/ Wang '24 \cite{wang2024codesigned}& Gangasani '24 \cite{gangasani202456gbps}/ Chen '25 \cite{chen2025optical}& Daudlin '25 \cite{daudlin2025nature}/ Wang '24 \cite{wang2024codesigned}\\
        \cmidrule{2-7}
        & \multicolumn{3}{r}{Total Power [W] / Area [mm$^2$]} & 449.68 / 428.60 & 1482.23 / 1395.36 & 456.13 / 431.95 \\
        \midrule
        \multirow{3}{*}{\textbf{(2) BW $\times$ 0.5}} 
        & 100 & 0.5--25 & 13--300 & Melek '26 \cite{melek2026terabit} / SuperCHIPS \cite{shih2021superchips}& Melek '26 \cite{melek2026terabit} / SuperCHIPS \cite{shih2021superchips}& Daudlin '25 \cite{daudlin2025nature}\\
        & 780 & 35--75 & 13--600 & Daudlin '25 \cite{daudlin2025nature}& Gangasani '24 \cite{gangasani202456gbps}& Daudlin '25 \cite{daudlin2025nature}\\
        \cmidrule{2-7}
        & \multicolumn{3}{r}{Total Power [W] / Area [mm$^2$]} & 154.32 / 156.83 & 440.96 / 479.48 & 156.62 / 158.02 \\
        \midrule
        \multirow{3}{*}{\textbf{(3) Dist $\times$ 2}} 
        & 20 & 1--50 & 52--1200 & Melek '26 \cite{melek2026terabit} & Melek '26 \cite{melek2026terabit} & Daudlin '25 \cite{daudlin2025nature}\\
        & 860 & 70--150 & 26--1200 & Daudlin '25 \cite{daudlin2025nature}/ Wang '24 \cite{wang2024codesigned}& Gangasani '24 \cite{gangasani202456gbps}/ Chen '25 \cite{chen2025optical}& Daudlin '25 \cite{daudlin2025nature}/ Wang '24 \cite{wang2024codesigned}\\
        \cmidrule{2-7}
        & \multicolumn{3}{r}{Total Power [W] / Area [mm$^2$]} & 316.37 / 309.35 & 3899.79 / 1602.47 & 318.08 / 313.16 \\
        \midrule
        \multirow{3}{*}{\textbf{(4) Dist $\times$ 0.25}} 
        & 20 & 0.125--6.25 & 26--600 & SuperCHIPS \cite{shih2021superchips}& SuperCHIPS \cite{shih2021superchips}& Daudlin '25 \cite{daudlin2025nature}/ Wang '24 \cite{wang2024codesigned}\\
        & 860 & 8.75--18.75 & 26--1200 & Melek '26 \cite{melek2026terabit} & Melek '26 \cite{melek2026terabit} & Daudlin '25 \cite{daudlin2025nature}/ Wang '24 \cite{wang2024codesigned}\\
        \cmidrule{2-7}
        & \multicolumn{3}{r}{Total Power [W] / Area [mm$^2$]} & 207.53 / 88.01 & 207.53 / 88.01 & 318.08 / 313.16 \\
        \bottomrule
    \end{tabular}
    }
\end{table*}

To evaluate our optimization framework, we compare it with a simplified baseline in Fig.~\ref{fig:full_results_comparison}. The Simple method selects links independently per net based only on reach constraints, choosing the feasible link with the highest peak bandwidth under shoreline limits. Although this greedy approach ensures feasibility, it can be inefficient in area and energy. As shown in Fig.~\ref{fig:full_results_comparison}, for Examples~1 and~2, the All-Link-Types strategy yields the same selections as the Electrical-Only case. This is due to short chiplet distances (0.5--6\,mm) and moderate bandwidth demands, where electrical links are more efficient. In contrast, Example~3 features longer distances (25--77\,mm) and higher bandwidth requirements, leading to a more demanding topology. Under these conditions, the optimizer selects optical links to improve both energy and area efficiency.

Table~\ref{tab:integrated_results_with_metrics} presents the results considering ECC. For Example~3, we perform case studies by scaling bandwidth and net distances. Case (1) (BW $\times$ 1.4) yields the same link selections as the baseline because the same high-bandwidth links remain required to satisfy the stricter bandwidth constraints. Case (2) (BW $\times$ 0.5) removes some selected optical links, as reduced bandwidth demands make electrical links sufficient. Case (3) (Dist $\times$ 2) increases reliance on optical or high-performance links. For example, Chen '25 \cite{chen2025optical} is selected in case (3) to handle longer distances. Case (4) (Dist $\times$ 0.25) favors low-power electrical links such as Melek '26 \cite{melek2026terabit} . Overall, these results show that the optimizer adapts link choices to minimize system-level power and area under varying constraints.

\section{Conclusion}
This paper presents a link-quality-aware pathfinding method that models ECC overhead for chiplet interconnects. Using RS$(86,K)$ FEC, CRC64, and Go-Back-N blocks, we quantify how ECC energy, goodput, and area/throughput vary with input BER under a delivered-BER target. Results show that ECC overhead can change link rankings and that CRC+ARQ can reduce RS strength—and decoder overhead—at moderate BERs while meeting stringent BER targets. We present a CP-SAT link assignment using ECC-corrected metrics with reach, bandwidth, and shoreline constraints, while leaving burst-error traces, BDP-aware ARQ throughput effects, latency, and routing congestion to future work.


\bibliographystyle{IEEEtran}
\bibliography{IEEEabrv,main}

\end{document}